# Probing the Relationship between Defects and Enhanced Mobility in MoS$_2$ Monolayers Grown by Mo Foil

*Sudipta Majumder[1], Vaibhav Walve[1], Rahul Chand[1], Gokul M.A.[1], Sooyeon Hwang[2], G. V. Pavan Kumar[1], Aparna Deshpande[1], Atikur Rahman[1]\**

[1]Department of Physics, Indian Institute of Science Education and Research, Pune, Maharashtra, 411008, India

[2]Center for Functional Nanomaterials (CFN), Brookhaven National Laboratory, US Department of Energy Office.

\*atikur@iiserpune.ac.in

## Abstract

In recent years, monolayer Molybdenum Disulfide (MoS$_2$) has emerged as a pivotal material in advancing next-generation optoelectronic technologies. Consequently, a pressing demand exists for the large-scale production of MoS$_2$ monolayers for fundamental research and commercial applications. Atomic vacancies, such as chalcogen vacancies, hold significant importance in changing the host material's electronic structure and transport properties. We present a straightforward one-step method for growing monolayer $MoS_2$ utilizing oxidized Molybdenum (Mo) foil using chemical vapour deposition and delve into the transport properties of as-grown samples. The controlled release of precursors from the Mo foil facilitates the growth of monolayer $MoS_2$ under a wide range of temperatures and growth time. Devices fabricated from these $MoS_2$ sheets exhibit excellent electrical responses, with the standout device achieving mobility exceeding 100 $cm^2V^{-1}s^{-1}$. Structural analysis and optical signatures unveiled the presence of chalcogen defects within these samples. To decipher the influence of inherent defects on the electronic transport properties, we measured low-temperature transport on two distinct sets of devices exhibiting relatively high or low mobilities. Combining the thermally activated transport model with quantum capacitance calculations, we have shown the existence of shallow states near the conduction band, likely attributed to sulfur vacancies within $MoS_2$. These vacancies are responsible for the hopping conduction of electrons in the device channel. Furthermore, our claims were substantiated through low-temperature scanning tunnelling microscopy measurements, which revealed an abundance of isolated and lateral double sulfur vacancies in Mo foil-grown samples. We found that these vacancies increase the density of states near the conduction band, inducing intrinsic n-type doping in the $MoS_2$ channel. Consequently, this elevated conductivity enhances the field-effect mobility of $MoS_2$ transistors. Our study offers insights into chalcogen vacancies in CVD-grown monolayer $MoS_2$ and highlights their beneficial impact on electronic transport properties.

## Introduction

Two-dimensional transition metal dichalcogenides (2D-TMDs) are an emerging class of materials and have attracted considerable attention due to their unique fundamental properties[1–7] and their potential applications in optoelectronic and nanoelectronic devices.[8–12] Among them,

Molybdenum Disulfide ($MoS_2$) is a widely studied 2D material for its numerous interesting properties such as direct band gap in the monolayer limit,[13] high Young's modulus,[14] and strong spin valley coupling.[15] These properties make $MoS_2$ a leading candidate for future Field Effect Transistors (FETs),[16,17] sensors,[18] photodetectors,[19] Light Emitting Devices (LEDs)[20] and photovoltaic devices.[21] For these reasons, the large-scale growth of monolayer $MoS_2$ (ML-$MoS_2$) is highly desirable.

Chemical vapour deposition (CVD) is a scalable, cost-effective technique to grow large-area ML $MoS_2$.[22,23] Generally, the mobility of $MoS_2$ monolayers grown by CVD is in the range of 0.8-30 $cm^2V^{-1}s^{-1}$,[24,25] much less than the theoretically predicted value of 410 $cm^2V^{-1}s^{-1}$.[26] The limited mobility observed in $MoS_2$ grown via CVD is commonly linked to the presence of sulfur vacancies, which occur due to their low formation energy.[27] Significant endeavours have been made to rectify these vacancies through various methods, aiming to restore intrinsic transport characteristics of $MoS_2$.[28,29] However, in certain instances, the defect repair process has led to a deterioration in device performance.[30–33] Moreover, the origin of ubiquitous n-type doping in monolayer $MoS_2$ is still not well understood. Some studies consider sulfur vacancies to act as donor states,[34–36] causing n-type doping and increasing channel mobility. Also, there are reports indicating that sulfur vacancies act as structural defects,[28,37] of which degrade the mobility of $MoS_2$ field effect transistors (FET) and that vacancies do not cause n-type doping.[38] Further, The uncontrolled supply of Mo precursors from $MoO_3$ powders can form oxides and other intermediate species along with $MoS_2$, altering the material's electrical and optical properties. Alternatively, Molybdenum foil (Mo foil) can also serve as a precursor for $MoS_2$ growth.[39–41] Generally, Oxygen ($O_2$) is used along with a carrier gas to activate the Mo foil surface by forming a Molybdenum Oxide ($MoO_x$) layer.[41] However, the presence of oxygen may corrode the samples or generate unwanted oxides in MoS$_2$, ultimately compromising the electrical properties of the devices. Electrochemical anodization has also been used to oxidized the Mo foil, but this technique primarily produces multilayered samples.[39]

In this study, we describe the synthesis of ML MoS$_2$ using Mo foil in a single-step CVD method and showed the sulfur vacancies formed naturally can be beneficial for the transport properties. Before the growth, the Mo foil was treated with oxygen plasma to create a thin layer of oxide, which acted as a precursor to the reaction. Oxygen gas was not introduced along with the carrier gas to oxidize the Mo foil, which prevents oxidation of MoS$_2$ during growth. To demonstrate the effectiveness of this method, ML MoS$_2$ was grown over a wide range of temperatures and for varying durations, which showed that the oxidized Mo foil provided a continuous supply of precursors throughout the temperature range and growth time. As-grown monolayer samples were characterized using X-ray photoelectron spectroscopy (XPS), Raman spectroscopy, Photoluminescence (PL) spectroscopy, and high-angle annular dark-field scanning transmission electron microscopy (HAADF-STEM) that revealed presence of sulfur defects in the sample. FETs were fabricated to investigate the effect of defects in the electrical properties of the ML-MoS$_2$. The devices showed a good electrical response, with the champion device having mobility exceeding 100 cm$^2$V$^{-1}$s$^{-1}$. We performed measurements in vacuum and low-temperature conditions to understand the effect of the environment and intrinsic defects on the transport properties of MoS$_2$. Our results showed a significant enhancement in mobility when placed in vacuum due to removed adsorbates. To understand the origin of n-type doping and the role of sulfur vacancies in the electrical transport of MoS$_2$ FETs, we studied two sets of devices with high and low mobilities. A thermally activated transport model revealed the presence of a finite density of states close to the conduction band of MoS$_2$, attributed to mid-gap states due to structural defects. These states enhance the conductivity and mobility of the devices by increasing the carrier concentration and screening the charged impurities in the channel. Below a particular temperature, the transport is governed by a variable range hopping transport. So, the enhanced density of states below the conduction band reduces the hopping distance between states and increases the mobility of the channel. To further establish our claim, we used low-temperature scanning tunnelling microscopy to characterize the nature of defects. Together with isolated single sulfur vacancies, we found an abundance of double sulfur vacancies in Mo foil-grown samples, increasing the density of states below the conduction band and shifting the Fermi level towards the conduction band, which enhanced n-type doping in ML $MoS_2$. Our study tried to shed light on the beneficial aspects of sulfur defects on the electronic properties of CVD-grown ML $MoS_2$.

**Results and Discussion**

We used an atmospheric pressure chemical vapour deposition system (APCVD) and applied a face-to-face metal precursor supply route to grow ML-$MoS_2$ flakes on $SiO_2$ substrates. To make the surface of Mo foil chemically active, we oxidized the surface with oxygen plasma. The details of the oxidation process are described in the Methods section. The $O_2$ plasma converted the exposed surface of the foil to $MoO_x$, which has a lower sublimation temperature (~400℃) than Mo metal (~2600℃). The XPS spectra of oxidized Mo foil in Figure 1b show two intense peaks at ~235.7 eV and ~232.6 eV, which could be assigned to the binding energies of Mo $3d_{3/2}$ and Mo $3d_{5/2}$ of the $Mo^{6+}$ oxidation state.[42] As previously reported, lower oxidation states ($MoO_{3-x}$) may give rise to two additional peaks in lower binding energies.[43] A less intense peak at ~228 eV could be attributed to the binding energy of Mo $3d_{5/2}$ of Mo metal.[44,45] The schematic of the APCVD arrangement is shown in Figure 1c. An alumina crucible containing the oxidized Mo foil was kept inside a quartz tube with a diameter of 3.5 cm. A clean $SiO_2$ substrate was kept on the alumina boat facing towards the Mo foil. Another crucible containing 500 mg sulfur was kept upstream. High-purity argon gas was used as carrier gas. The growth was done as follows: the tube was purged with argon gas with 500 sccm for 30 min to get rid of any moisture and oxygen present inside the tube, then the flow rate of the carrier gas was decreased to 30 sccm, and the furnace was ramped at a rate of 300°C / hour to reach the growth temperature, which was maintained constant throughout the reaction time. After completion of the reaction, the furnace was left to cool down naturally. Oxidized Mo foil provides a controlled supply of precursor over a wide range of temperatures. The evolution of $MoS_2$ in the temperature range from 600°C to 800°C is shown in the Supplementary Figures S2a-S2d. We observed variations in grain size and nucleation density at different temperatures. At the growth temperature of 600°C, the sublimation rate of precursors is relatively low, so the size of grains on the substrate is small with a large number of nucleation points (Figure S3a). As the temperature increased, the rate of diffusion of precursors increased, resulting in larger domain sizes and fewer nucleation points. We didn't observe any significant multilayer growth up to 800°C. Further, we explored the effect of growth time on the morphology of $MoS_2$ monolayers. Supplementary Figures S2e-S2g show that the domain size of $MoS_2$ triangles increases with increasing growth duration for a fixed temperature. The increasing lateral growth of $MoS_2$ domains with time can be due to the low concentration of $MoO_{3-x}$ supply from the Mo foil surface, which helps the 2D planer growth of $MoS_2$.[46] The significantly low density of multilayers indicates that the oxidized Mo foil-assisted growth has exceptionally high control over the precursor than the growth with powdered $MoO_3$ as a Mo precursor source.

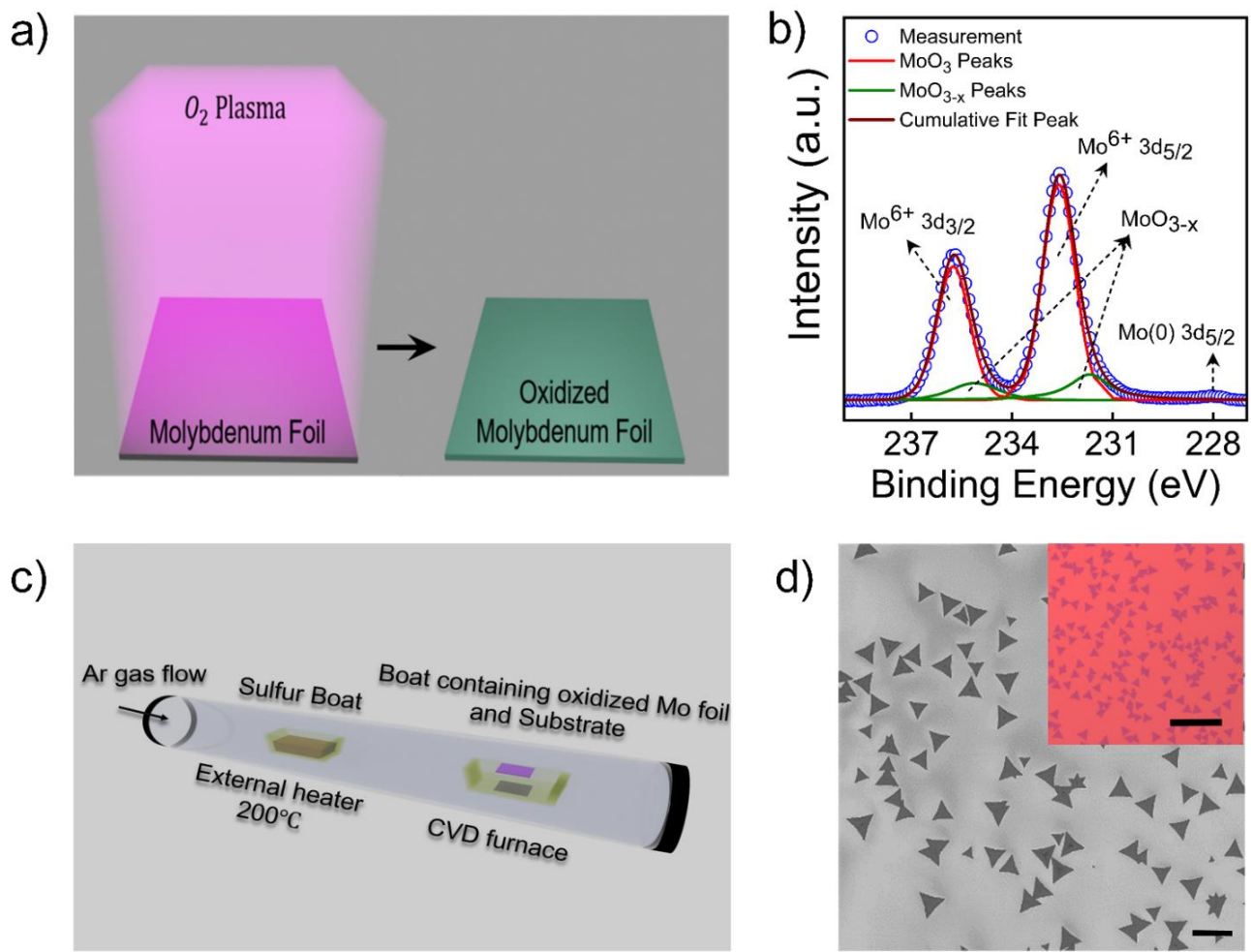

**Figure 1** APCVD growth of Mo foil assisted MoS$_2$ monolayers. (a) Oxidation process of the Mo foil by high-power oxygen plasma before the growth. (b) XPS sprectra of oxidized Mo foil. Two intense peaks of Mo$^{6+}$ oxidation peaks of Molybdenum Trioxide (MoO$_3$) are denoted with red line. Two additional peaks at lower binding energies can be attributed to lower oxidation states of Molybdenum Oxides (MoO$_{3-x}$). A very weak intensity peak can be found at ~228 eV which is assigned to Mo 3d$_{5/2}$ peak of metallic Molybdenum (c) Schematic of face-to-face metal precursor supply route in APCVD system. (d) SEM images of as-grown MoS2 triangles on SiO$_2$ substrate. Scale bar is 40 μm. An optical image is shown in the inset. The scale bar is 200 μm

The structural characterization of as-grown ML-MoS$_2$ was thoroughly done. Figure 2a displays XPS data of as-grown MoS$_2$, illustrating the presence of Mo$^{4+}$ peaks at 232.9 eV and 229.7 eV for the Mo 3d$_{3/2}$ and Mo 3d$_{5/2}$ peaks, respectively. These are the binding energies of Mo 3d doublet corresponding to the intrinsic MoS$_2$ (i-MoS$_2$). Another peak at 226.9 eV is evident, corresponding to the 2s peak of S. These binding energy values agree well with the XPS data of the 2H phase of MoS$_2$ previously reported in the literature.[47] Two additional peaks appear at lower binding energies than the Mo 3d doublet of i-MoS$_2$.[48] These can be assigned to the non-stoichiometric MoS$_x$ (0<x<2), indicating sulfur defects in CVD-grown MoS$_2$. Further, we did not observe any peaks due to oxidation of the samples at higher binding energies. This confirms that no interstitial oxygen is present in MoS$_2$. To determine the quality and presence of defects in the CVD-grown MoS$_2$ monolayer, Raman and PL spectra of the as-grown samples were examined. In Figure 2b, the Raman peaks were modelled with a Lorentz function. Two peaks which appeared at 383.3 cm$^{-1}$ and 402.2 cm$^{-1}$, correspond to E$^1_{2g}$ and A$_{1g}$ peaks of MoS$_2$, respectively. The difference between the peaks was found to be 18.9 cm$^{-1}$, indicating that MoS$_2$ is monolayer.[49,50] The shoulder peak at a slightly lower energy of ~377 cm$^{-1}$ arises due to the disorder in MoS$_2$ and can be assigned to LO (M) mode.[51] In Figure 2d, the PL spectrum of as-grown flakes was fitted with a Gaussian-Lorentzian cross-function, which yields two peaks at 1.87 eV and 2.02 eV, corresponding to A and B excitonic peaks, respectively, arising from spin-orbit splitting in the valance band.[52] AFM confirmed the monolayer nature of the MoS$_2$ flakes. Supplementary Figure S3 shows the height profile yields a thickness of 0.63 nm, typical of monolayers of MoS$_2$.[53–55] To dig deeper into the atomic structure of MoS$_2$ monolayers, we used high-angle annular dark-field scanning transmission electron microscopy (HAADF-STEM). Figure 2e exhibits a STEM image of ML-MoS$_2$. We found sulfur vacancies present in ML-MoS$_2$, which are marked by red arrows. Note that these vacancies can also arise due to electron irradiation. Hence, we have explored the sulfur vacancies present in the samples by scanning tunnelling microscopy (STM), which will be discussed later in the manuscript. We did not observe any molybdenum vacancies in the lattice. The diffraction pattern of the selected area in Figure 2f provides evidence for the single crystalline nature of the MoS$_2$ lattice and hexagonal symmetry.[56] IFFT image of the atomic planes of MoS$_2$ (see inset of Figure 2f) revealed the interplanar distance is 0.27 nm, which corresponds to the (100) planes of $MoS_2$.[57]

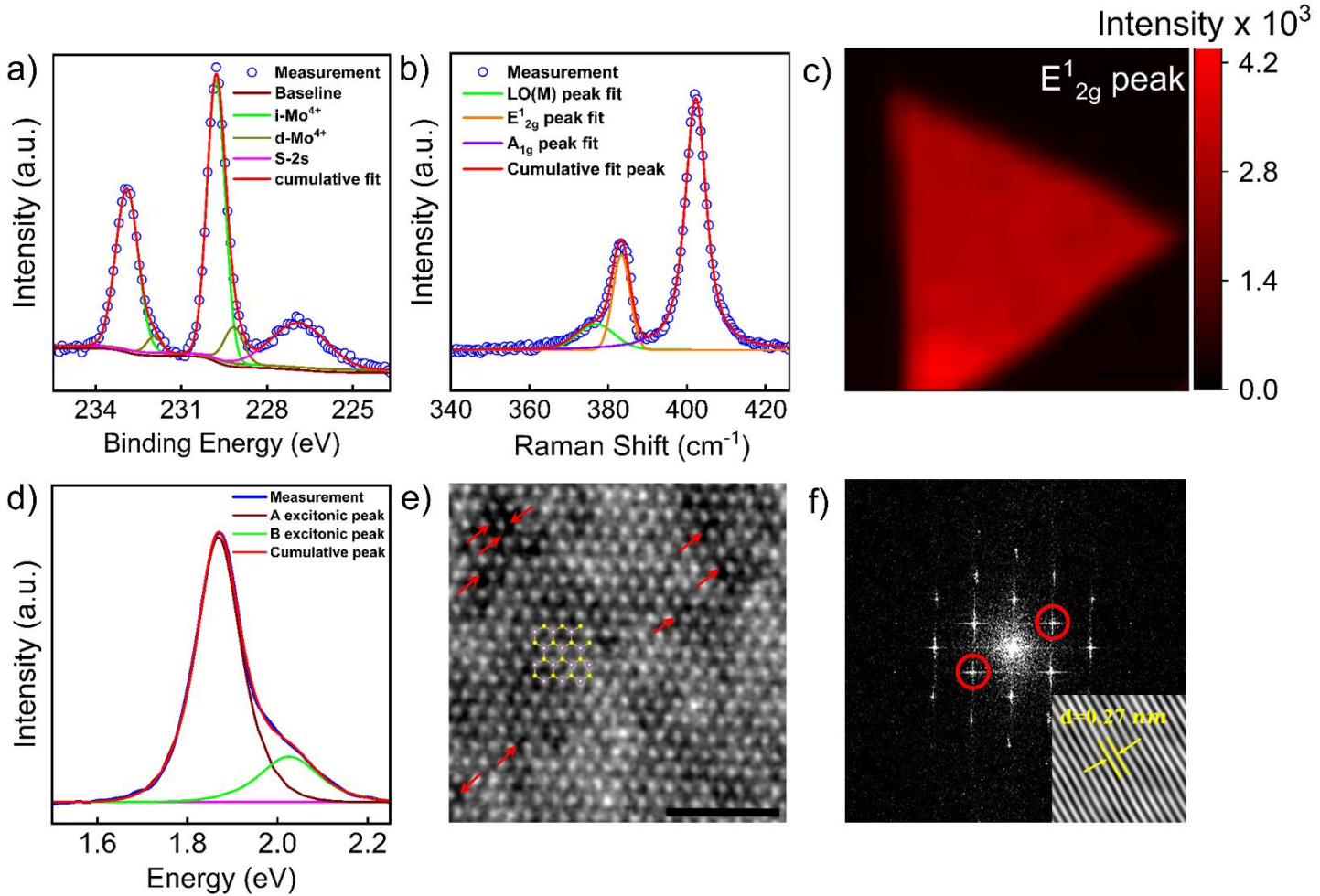

**Figure 2** Structural characterization of Mo foil grown $MoS_2$ a). XPS spectra of Mo 3d core level of as-grown $MoS_2$. The pristine $MoS_2$ shows a characteristic doublet corresponding to $Mo^{4+}$ state (i-$Mo^{4+}$) and S 2s state. A less intense doublet is found in the lower binding energies corresponding to the nonstoichiometric $MoS_x$ (0<x<2), indicating sulfur defects present in the CVD-grown sample. b) Raman spectra of $MoS_2$. The separation between $E^1_{2g}$ and $A_{1g}$ peak is 18.9 $cm^{-1}$, indicating monolayer nature. A shoulder peak at the left of $E^1_{2g}$ mode arises due to the defects denoted by LO(M) mode. c) Raman intensity map of $E^1_{2g}$ peak. d) Photoluminisense of monolayer $MoS_2$. Two distinct peaks are present in the spectrum, A excitonic peak and B excitonic peak. e) High resolution aberration corrected TEM image showing atomic structure of monolayer $MoS_2$. Red arrows indicates the position of sulfur vacancies in CVD grown sample. f) Diffraction (FFT) image of selected area showing hexagonal symmetry of $MoS_2$ lattice. The Bottom-Right image shows the planes corresponding to (100) planes.

We further explored the electrical performance of Mo foil-grown ML MoS$_2$ through electrical measurements. The output (I$_d$-V$_d$) and transfer characteristics (I$_d$-V$_g$) were measured between the source (S) and drain (D) contacts of a device having a channel length of 15 μm and width of 8 μm. Figure 3a shows a representative optical image of a typical device. The output characteristics show (Figure 3b) channel resistance modulation with increasing gate voltage. The slight nonlinearity in $I_d$-$V_d$ characteristics could result from Schottky contact formed at a metal-semiconductor junction. The full gate modulation of channel resistance in ambient conditions is shown in transfer characteristics in Figure 3c at a fixed source-drain bias ($V_d$) of 1 V in both linear and logarithmic scales. The ML MoS$_2$ FET behaves as an n-type channel.[58] The field effect mobility can be calculated from the following expression $\mu_{F.E.} = \frac{L}{C_{ox}W} \frac{dI_d}{dV_g} \frac{1}{V_d}$ where L and W are the channel length and width, respectively, $C_{ox} = 1.2 \times 10^{-8}$ F cm$^{-2}$ is the capacitance per unit area of a 285 nm thick SiO$_2$ gate dielectric and $\frac{dI_d}{dV_g}$ is the transconductance of the FET device calculated from the linear part of the transfer characteristics.[53] The mobility of ML MoS$_2$ FETs under ambient conditions was calculated without any passivation of defects or device annealing to see the effect of adsorbates and intrinsic defects, as device annealing may create additional defects in the MoS$_2$ channel.[59]

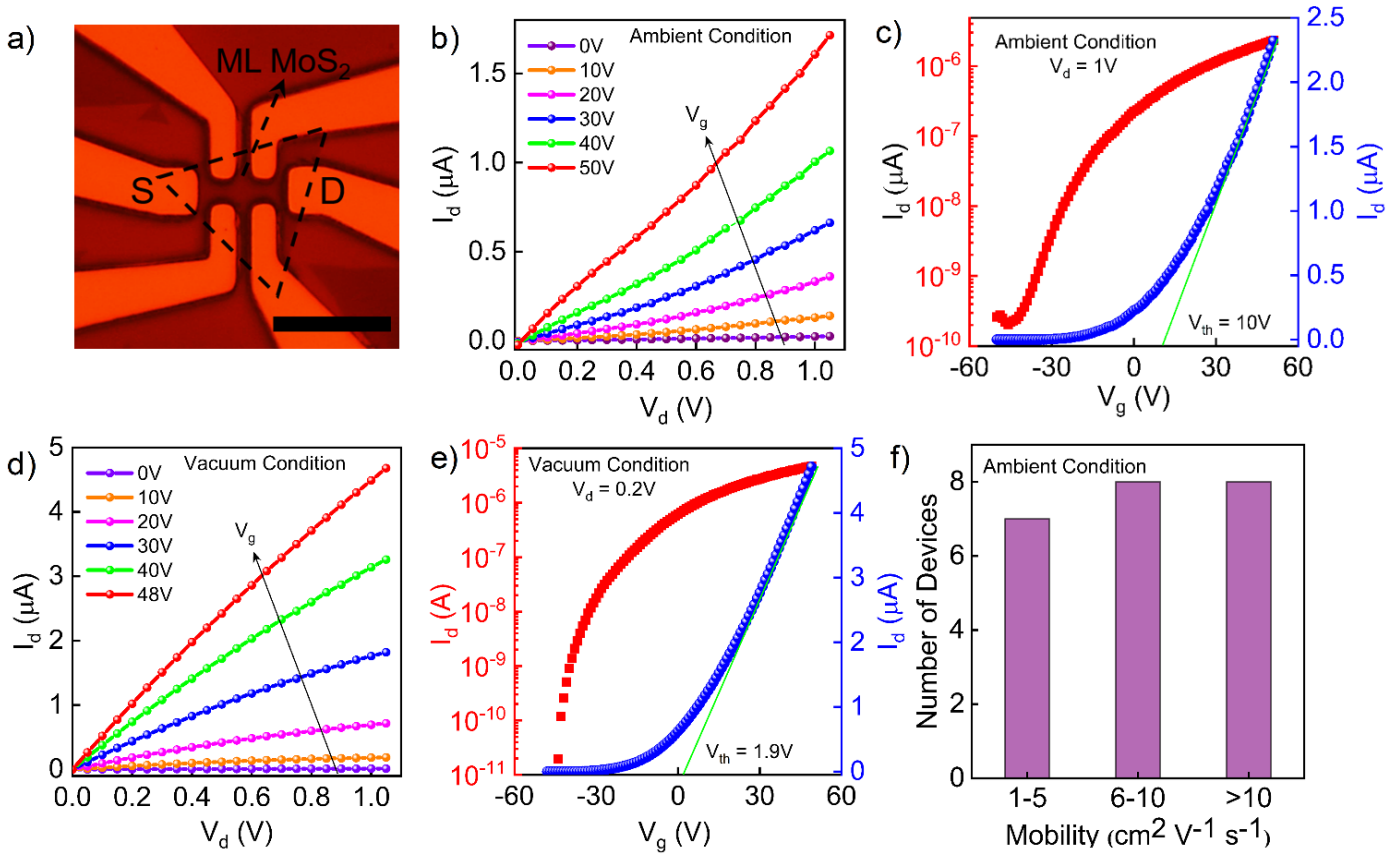

**Figure 3**. Fabrication and characterization of ML MoS$_2$ field effect transistors (a) Optical image of a typical MoS$_2$ FET with source and drain contact maked. The scale bar is 20 μm. (b) Output characteristics. i.e., source to drain current (I$_d$) versus bias voltage (V$_d$) at different back gate voltages (V$_g$) in ambient condition a device having mobility of 11 cm$^2$V$^{-1}$s$^{-1}$ . (c) Transfer characteristics, i.e., source to drain (I$_d$) characterestics versus gate voltage (V$_g$) at a fixed bias voltage of 1 V . The threshold voltage (V$_{th}$) is calculated using linear extrapolation method. (d) Output characteristics the device in vacuum condition showing increase channel conductance in absence of adsorbates. (e) Transfer characteristics of the device in vacuum at a source-drain voltage of 0.2 V. The conductivity is increased and threshold is shifted to more negative side, indicating increased doping in channel in vacuum condition in absence of adsorbates. (f) The statistics of field effect mobility of Mo foil grown ML MoS$_2$ devices calculated for different devices in ambient conditions.

Owing to its large surface-to-volume ratio, the transport properties of ML-MoS$_2$ are susceptible to environmental conditions.[60] To observe the effect of the environment on device performance, we measured the output and transfer characteristics of devices in vacuum. We found that in vacuum conditions (base pressure of 10$^{-6}$ mbar), the mobility of the devices increased significantly. The highest mobility observed was ~101 cm$^2$V$^{-1}$s$^{-1}$, which was a significant increase from its initial value of 11 cm$^2$V$^{-1}$s$^{-1}$, measured in ambient conditions. Figures 3d and 3e show the device's output and transfer characteristics, respectively. The output characteristics curves became linear in the vacuum condition, indicating ohmic contacts. These observations of increased channel conductance and improved contact resistance in vacuum have been previously reported and are primarily attributed to the physisorbed H$_2$O and O$_2$ molecules on the MoS$_2$ surface.[60,61] These molecules bind strongly with the sulfur vacancies on the surface of CVD-grown $MoS_2$, and due to charge transfer between the $MoS_2$ channel and adsorbed molecules, the channel conductance decreases.[62] Moreover, these trapped molecules act as Coulomb scattering centres, limiting the mobility of 2D TMDs.[63] The threshold voltage ($V_{th}$) was determined from the linear region of the $I_d$-$V_g$ graph by linear extrapolation method, as shown in Figures 3c and 3e. We observed that the device's $V_{th}$ drops to 1.9V in the vacuum from an ambient value of 10V due to the increased doping of the $MoS_2$ channel in the absence of adsorbates. Figure 3f displays the histogram of the mobility values and the number of devices in ambient conditions. The gate hysteresis loop of FETs is also highly sensitive to environmental effects.[64] We measured the gate hysteresis loop in ambient and vacuum conditions. As the sweep rate of gate voltage also affects the hysteresis, we did all measurements under the constant sweep rate of 0.2V/s. Figure S4 (see Supplementary Materials) shows hysteresis loop width decreased significantly in vacuum conditions (~$10^{-6}$ mbar). The remnant hysteresis can be due to the charge trapping and de-trapping in the trap states between the $MoS_2$ and $SiO_2$ interface and intrinsic defects inside the $MoS_2$ channel.[65,66] As the temperature drops, the width of the hysteresis also nearly vanishes. The capture and release of charge carriers in the interfacial trap states are known to be thermally activated. So, as we reduced the temperature, the density of trapped electrons also decreased.

It is essential to mention that the mobilities of Mo foil-grown samples are generally higher compared to samples grown using other precursors in CVD. To understand the reason behind higher mobility, we studied the temperature (T) dependent behaviour of mobility (µ). Specifically, we analyzed two sets of devices: one with relatively high mobility (HM) and the other with relatively low mobility (LM) having mobility values >100 $cm^2V^{-1}s^{-1}$ and <50 $cm^2V^{-1}s^{-1}$, respectively, at 300K in vacuum. Figure 4a shows the temperature-dependent mobility value of LM and HM devices from 300K to 11K. The mobility of charge carriers in 2D semiconductors is affected by Coulomb Impurities (CI) acoustic and optical phonon scatterings. The resultant mobility in 2D semiconductors can be modelled with Matthiessen's rule, $\frac{1}{\mu} = \frac{1}{\mu_{ph}} + \frac{1}{\mu_{CI}}$, where $\mu_{ph}$ is the phonon limited mobility, and $\mu_{CI}$ is Coulomb impurity limited mobility.[67] For HM device at lower temperatures, mobility decreased with decreasing temperature due to the CI limited transport in the channel and can be fitted with $\mu_{CI} \sim T^{-\alpha}$. The value of α for the HM device was calculated to be 0.5, which differs from the predicted value of 1.5.[68] As the temperature rose above 230K, the mobility of the H.M. device decreased, indicating electron-phonon scattering in the H.M. device. In this region, mobility is fitted with $\mu_{ph} \sim T^{-\beta}$. The value of β was calculated to be 0.3. Notably, this value also differs from the predicted value of 1.69 in $MoS_2$.[69] On the other hand, The mobility of the LM device consistently increases with increasing temperature over the entire temperature range, suggesting that its mobility is solely limited by CI. To qualitatively see the effects of various mobility limiting factors in both devices, we calculated electron density (n) and density of interfacial trap states ($D_{it}$) between the $MoS_2/SiO_2$ interface. At a fixed gate voltage of $V_g$ = 48 V, the concentration of electrons in the HM device is $\sim 3.53 \times 10^{12}$ $cm^{-2}$ and L.M. device has a concentration of $\sim 2.47 \times 10^{12}$ $cm^{-2}$ at 300K, calculated using the parallel plate capacitor model where n = $\frac{C_{ox}(V_g - V_{th})}{q}$.[17] We can calculate the density of interfacial trap states from the devices' subthreshold swing (S.S.), which is given by the $D_{it} = (\frac{q(SS)}{kT\ln(10)} - 1) \times \frac{C_{ox}}{q}$. In the subthreshold region, the value of S.S. (SS=$\frac{dV_g}{d(\log I_d)}$) is 4V/decade for LM device and 3V/decade for HM device. $D_{it}$ is found to $4.96 \times 10^{12}$ $eV^{-1}$ $cm^{-2}$ for LM device and $3.7 \times 10^{12}$ $eV^{-1}$ $cm^{-2}$ for HM device. As the impurity density limits the mobility of atomically thin materials at low temperatures, the

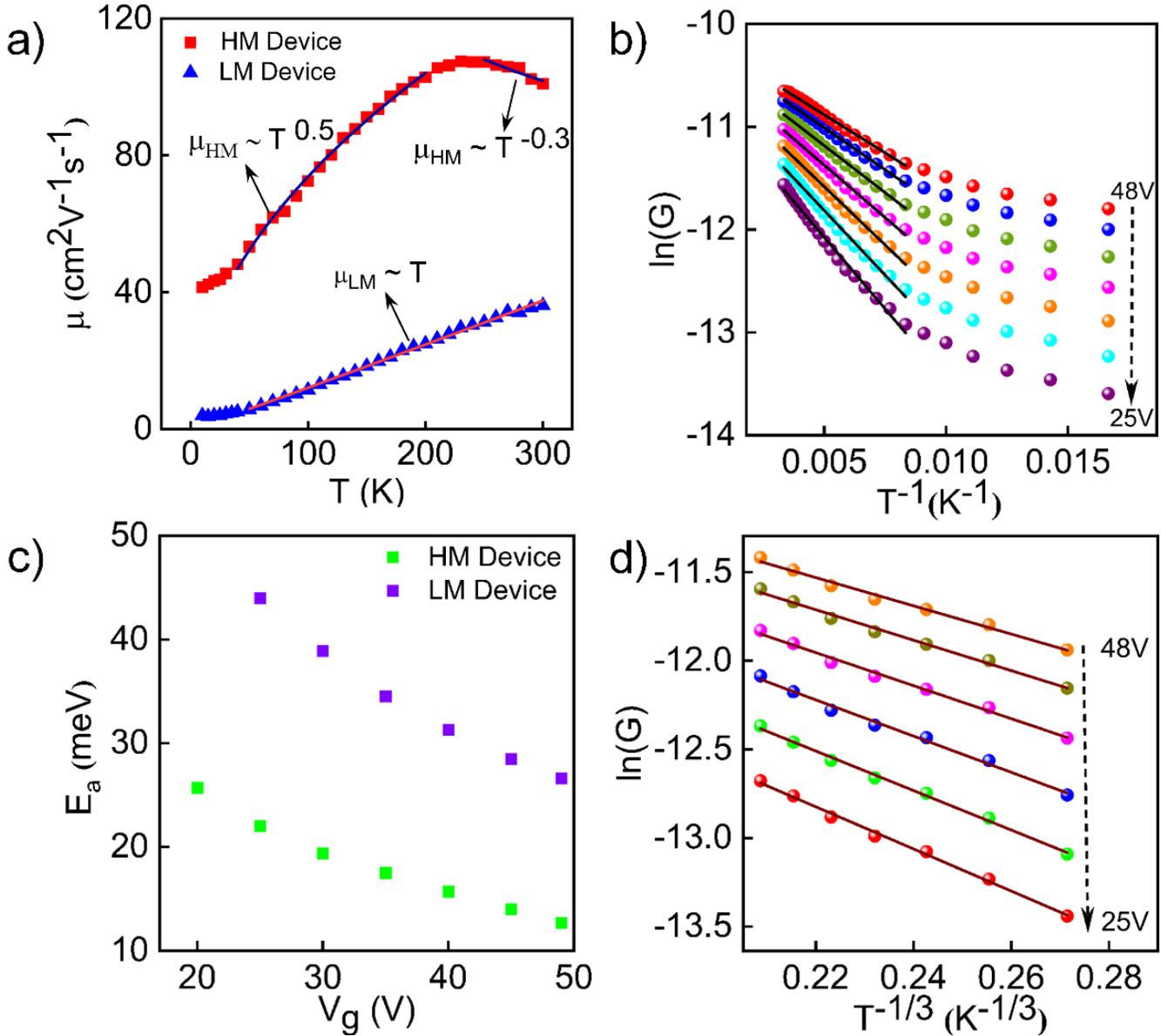

**Figure 4**. Low temperature transport in Mo foil grown MoS$_2$ FETs (a) Field effect mobility (μ) of low mobility (LM) and high mobility (HM) devices as a function of temperature (T). Two distinct regions can be seen in HM device inducating both coulomb impurity limited transport and phonon scattering in higher temperatures. On the other hand in LM device mobility is only limited by coulomb scattering. (b) logarithm of conductivity ln(G) as a function of inverse of temperature T$^{-1}$ of HM device plotted for different gate voltages from 48V to 25V. The high temperature region (>125K) is fitted with thermally activated transport model. (c) The activation energy (E$_a$) of LM and HM device as a function of gate voltage. The decrease of activation energy with increasing gate voltage indicates the fermi level approaches closer to the conduction band. (d) In lower temperatures than 125K the conduction can be described by Mott variable range hopping (VRH) conduction. The logarithm of conductivity, ln(G), is plotted with T$^{-1/3}$ for dirrerent gate voltages.

Coulomb scattering is more effective in LM devices having more interfacial defect states.[70] But the HM device having higher electron density in the channel more effectively screens the effect of long range charge impurities from the substrate, thus having a higher mobility at room temperature.[71] With reduced Coulomb scattering and fewer interfacial trap states, the phonon scattering mechanisms can dominate the charge transport in the HM devices near room temperature.

To further explore the effect of intrinsic disorders and the impact of vacancies from the temperature-dependent transport, we have plotted the natural logarithm of conductivity (ln(G)) as a function of the inverse of temperature ($T^{-1}$) for various gate voltages (see Figure 4b). Two distinct regions were observed in the plot. At high temperatures (>125K), the graph was fitted with a thermally activated transport model $G = G_0 \exp\left(-\frac{E_a}{kT}\right)$, where $E_a$ is the activation energy, k is the Boltzmann constant.[72] From the ln(G) vs $T^{-1}$ plot Figure 4b, we have calculated the activation energy for various gate voltages for both HM and LM devices. As seen in Figure 4c, the activation energy decreases with increasing gate voltage because the gap between the fermi level ($E_F$) and the edge of the conduction band ($E_C$) reduces as we increase gate voltage ($E_a = E_C - E_F$). The activation energy of LM device is higher than HM device at all gate voltages. At high gate voltages ($V_g$ = 48V), the activation energies in LM and HM devices are 26.7 meV and 12.8 meV, respectively. These values are close to the recently calculated activation energy for exfoliated $MoS_2$ monolayers[73] and are lower than the activation energy (96 meV) of the interfacial trap states.[74] The low activation energy indicates that thermally activated transport in these devices is likely due to the donor states arising from sulfur vacancies. Further, the activated transport is not due to the Schottky barrier, as the expected barrier height between $MoS_2$ and Cr is ~ 0.46 eV in vacuum.[61] So, the activated transport comes from the channel itself. The enhanced doping in HM device shifts the fermi level towards the conduction band edge, and consequently, HM device has lower thermal activation energy than LM device.[75] This is also evident from the observation that the threshold voltage of the HM device is more on the negative side (Supplementary Figure S8).

We calculated the density of states, N($E_F$), at the Fermi level to further verify our claim. The density of states can be calculated from the gate dependence of activated energy previously calculated using a thermally activated transport model and quantum capacitance ($C_d$) expressed as $\frac{dE_F}{dV_g} = -\frac{dE_a}{dV_g} = \frac{eC_{ox}}{C_{ox}+C_d}$, where $C_d = e^2 N(E_F)$.[76] At a higher gate voltage range ($V_g$ = 48V), The quantum capacitances are calculated to be 25.5 µF cm$^{-2}$ and 36 µF cm$^{-2}$ of LM and HM devices, respectively. The density of states is estimated to be $2.27 \times 10^{14}$ eV$^{-1}$cm$^{-2}$ and $1.43 \times 10^{14}$ eV$^{-1}$cm$^{-2}$ for HM and LM devices. These values are close to reported values for the density of states due to sulfur vacancies near the conduction band edge.[73,77] Notably, the density of states at high gate voltages is two orders greater than the interfacial states, which is of the order of ~ $10^{12}$ eV$^{-1}$ cm$^{-2}$, indicating that the interfacial trap states do not significantly contribute to the overall density of states. The HM device has a higher density of states near the Fermi level, providing more electrons for conduction and enhancing the channel conductivity. The mobility variation in these devices can also be described by the variable range hopping (VRH) conduction in the low-temperature regime (<125 K). We found that in low temperatures, the conductivity of the channel shows a $T^{-1/3}$ dependence (Figure 4d), suggesting 2D Mott VRH conduction.[78,79] In MoS$_2$, the hopping of electrons can take place through the sulfur vacancy sites. If there are more sulfur vacancies, the effective distance between hopping locations will decrease, increasing the hopping probability, which can increase the mobility of the channel.[32] So, the HM device with a higher density of sulfur vacancies should also have higher mobility.

The presence of defects in both LM and HM devices was confirmed by Raman and PL spectroscopy. Supplementary Figure S8 shows the Raman spectrum of both devices. The FWHM and peak positions of E$^1_{2g}$ and A$_{1g}$ peaks for HM and LM devices are given in the table below. The FWHM of both the peaks of Raman, $E^1_{2g}$ and $A_{1g}$, is higher for HM devices, indicating that the defect concentration is higher in devices with higher mobility devices.[80]

**Table 1.** Comparison of Raman and PL data for HM and LM device

| Device | $E^1_{2g}$ Peak ($cm^{-1}$) | $A_{1g}$ Peak ($cm^{-1}$) | $E^1_{2g}$ FWHM ($cm^{-1}$) | $A_{1g}$ FWHM ($cm^{-1}$) |
|---|---|---|---|---|
| HM | 383.2 | 402 | 6 | 6.8 |
| LM | 383.2 | 403 | 5.1 | 6.1 |

Additionally, the $A_{1g}$ peak is redshifted by 1 cm$^{-1}$ in the HM device, where the $E^1_{2g}$ peak is unchanged. This can be explained by the tight coupling of $A_{1g}$ phonons with electrons.[81,82] The higher doping in the HM device caused the red shift in $A_{1g}$ peak. The PL intensity of HM device is less than that of LM device, as shown in Supplementary Figure S9. The reduction and broadening in PL intensity can be described by the formation of tightly bound trions due to higher doping in HM device.[82,83] Additionally, the B/A excitonic peak ratio is higher for HM device, which is evident from the figure. This can be due to a higher density of defects in HM device.[84] Elevated electron density can substantially increase the population of B excitations through non-radiative recombination pathways.[85] This phenomenon could result in an increased density of B excitons within a high-mobility device.

To further confirm our claim and explore the detailed nature of the defects in Mo foil-grown MoS$_2$ samples, we have performed low-temperature STM measurements. In the STM topography, we have successfully identified ML MoS$_2$ flakes on Highly Ordered Pyrolytic Graphite (HOPG). The corresponding large-area STM topography is shown in Supplementary Figure S10. Atomic-scale imaging has confirmed a higher density of sulfur vacancies in various regions of the samples, typically ranging from $10^{13}$ to $1.6 \times 10^{13}$ per centimetre square, as depicted in Figure 5a. Notably, we have identified single vacancies ($S_1$), double vacancies ($S_2$), and multiple sulfur vacancy defects ($S_n$). These defects exhibit lateral orientation, contrasting with the vertical double or multiple defects reported previously.[86] We have distinguished $S_1$ and $S_2$ based on the shape of the defects in the STM topography, as shown in Figure S11. The $S_1$ defect preserves the triangular symmetry in the contrast, while the $S_2$ defect shows an hourglass-like shape contrast in the STM topography. Line scan profiling across the surface has revealed a lattice spacing of 0.32 nm. We observed that the depth over the single ($S_1$) and double sulfur ($S_2$) vacancies are nearly the same, around 100 pm, confirming that both are lateral defects and no sulfur is missing from the downward direction. The multiple sulfur vacancies show a depth of around 140 pm, possibly due to a Mo vacancy or an underlying S atom missing. As the presence of Mo defects was not evident in our HAADF-TEM images, we conclude that it is more likely to be the effect of underlying S vacancies. Furthermore, we observed an irregular arrangement of sulfur atoms around double or multiple vacancies, suggesting localized lattice strain within the structure. The atomic spacing was also observed to be modified, ranging from 0.3 nm to 0.32 nm. Scanning tunnelling spectroscopy (STS) was conducted on both single and double sulfur vacancy defects, as well as in pristine regions, revealing a band gap for all the areas. The pristine sample shows a band gap of nearly 1.8 eV, close to ML-MoS$_2$. The pristine region is slightly n-doped, as depicted in Figure 5b. The reduced band gap of the pristine region and shift of Fermi level towards the conduction band can be due to the sulfur defects of the surroundings.[87] The height profile over the defect is shown in Figure S11, which matches well with ML MoS$_2$. Spectroscopy over the isolated single defect ($S_1$) and lateral double sulfur defect ($S_2$) shows a lesser bandgap of 1.7 eV, with a shift in the Fermi energy closer to the conduction band, as clearly visible in the spectroscopy shown in Figure 5b. This can be attributed to donor states close to the conduction band due to sulfur vacancies.[88] Additionally, spectroscopy over the $S_1$ and $S_2$ defects shows additional midgap states occurring in the band gap, as observed in previous reports, which was previously attributed to the tip-induced band bending.[88] Figure 5d shows the density of states mapping along 1 to 9 positions of Figure 5c. We have identified defect states in the valence band for both types of defects and an enhancement of the states for double sulfur vacancies. Interestingly, this additional density of states shifted the conduction band edge by 100 meV. This increased density of states near the conduction band minima (CBM) may explain the higher conductivity observed in these samples. Our measurements were conducted at lower temperatures, where charge carriers could become trapped in these defect states. Our transport measurements supported this scenario, demonstrating lower conductivity in the samples at lower temperatures. As the temperature increased, thermal energy removed the barrier imposed by trapped states, allowing charge carriers to move freely between defect states, resulting in higher conductivity in the samples.

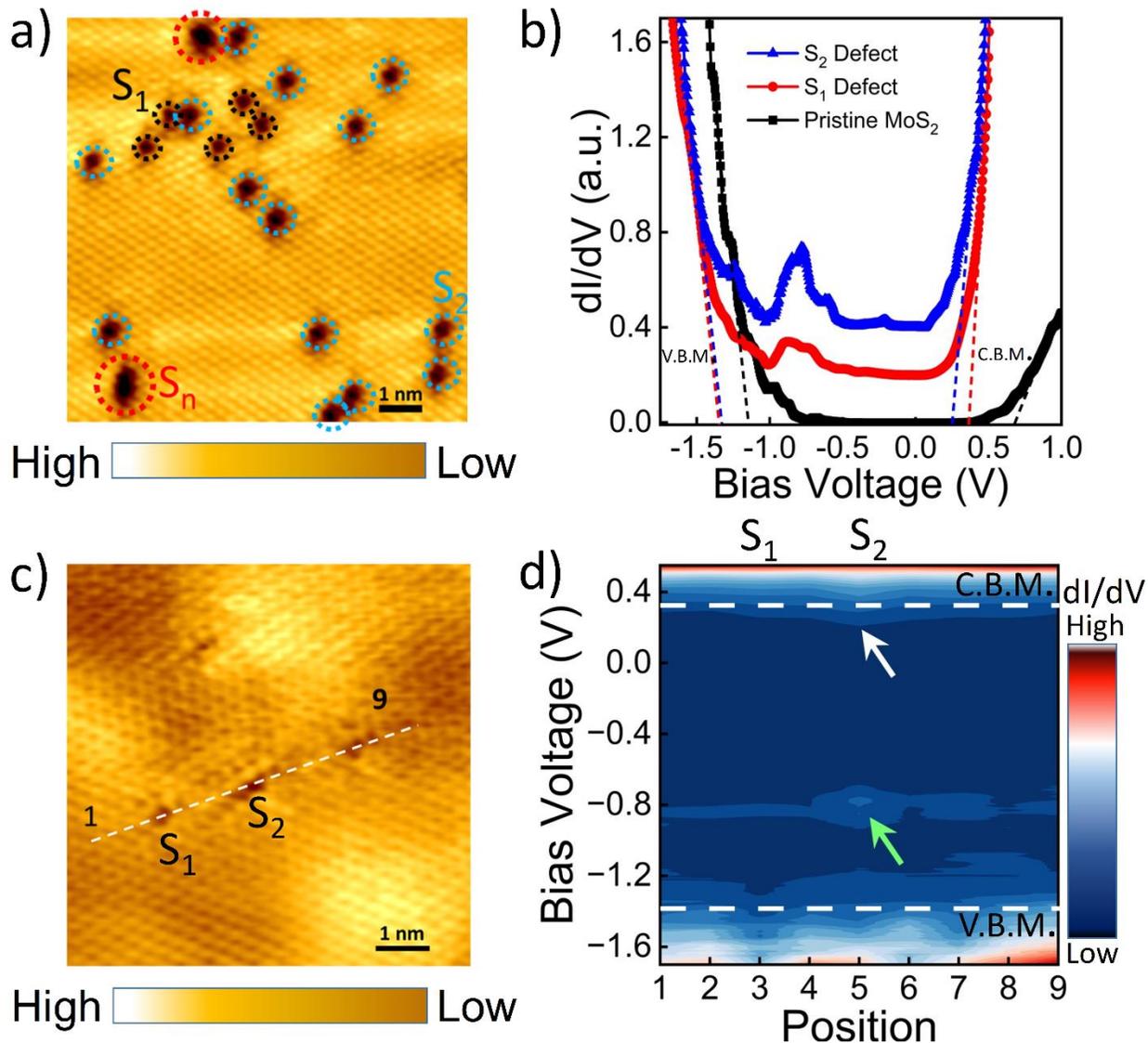

**Figure 5** STM measurements on Mo foil grown $MoS_2$ and characterization of different tyoes of sulfur defects present in the sample a) STM topography with identification of the defects. (VB=1V, $I_s$ = 150pA) b) Point spectroscopy over the region away from the all defects (Pristine $MoS_2$), single sulfur vacancy defect ($S_1$) and double sulfur vacancy defect ($S_2$). c) $S_1$ and $S_2$ defects shown in the STM topography. (VB=0.6V, Is=500pA) d) Position dependent spectroscopy over the $S_1$ and $S_2$ defects shown in the (c). DOS on $S_2$ defect increased at C.B.M. shown by white arrow and appearance of mid-gap defect states shown by the green arrow.

**Conclusion**

In summary, we report a one-step process to grow monolayer $MoS_2$ using oxidized Mo foil. The controlled supply of precursors allows the growth of $MoS_2$ in a wide range of temperatures and time. The devices show good electrical response and mobility. The effect of temperature and ambient conditions were explored. The presence of defects in these samples was thoroughly explored by XPS, Raman, Pl and HAADF-STEM. Electrical transport studies were conducted further to understand the effect of defects in these samples. Temperature-dependent transport studies in two devices with different mobilities revealed that comparatively high-mobility devices have a higher density of donor states close to the conduction band, likely due to the sulfur vacancies formed during the growth process. Low-temperature STM and STS measurements were conducted to support the findings from the transport experiments and further understand the defects in ML $MoS_2$. We found an abundance of double sulfur vacancies in these samples and isolated double sulfur vacancies, both increasing the density of electrons near the conduction band and shifting the fermi level towards the edge of the conduction band, which goes well with our findings with electrical measurements. Overall, our results demonstrate a highly effective method for growing ML-$MoS_2$ and provide a thorough study of the defects present in these samples and their effect on the electronic transport and band structure of ML MoS$_2$. Our analysis reveals by tuning the concentration of sulfur vacancies, desired electrical and optoelectronic properties can be obtained.

## Methods

**Oxidization of Molybdenum Foil.** A Molybdenum (Mo) foil of length 15 mm, width 6.5 mm and 1.0 mm thickness (Sigma-Aldrich, 99.9 % purity) was cleaned with acetone and IPA under ultrasonication for 10 min. The foils were then oxidized using oxygen plasma having 0.4 mbar pressure at 60-watt power in a diener Zepto plasma cleaner for 10 mins. The oxidation time and power of foils were optimized to get good coverage of the ML MoS$_2$ on the substrate. The foil can be reused after the growth. To reuse the foil, 800 grit sandpaper was used to polish it well, followed by acetone and IPA cleaning.

**CVD growth of MoS$_2$.** Monolayers of MoS$_2$ were synthesized in an atmospheric pressure chemical vapour deposition (APCVD). An alumina crucible containing the oxidized Mo foil was kept inside a quartz tube with a diameter of 3.5 cm. Degenerately doped Si substrates with 285 nm SiO$_2$ coating were cleaned with acetone and I.P.A. under ultrasonication followed by oxygen plasma treatment and kept on the alumina crucible with the SiO$_2$ side facing the Mo foil. Another crucible containing 500 mg sulfur was kept upstream. High-purity argon was used as carrier gas. The growth was done as follows: the tube was purged with argon gas with 500 sccm for 30 min to get rid of any moisture and oxygen present inside the tube, then the flow rate of carrier was decreased to 30 sccm, and the furnace was ramped at a rate of 300°C / hour to reach the growth temperature, which was maintained constant throughout the reaction time. After completion of reaction, the furnace was left to cool down naturally.

**Optical and structural characterization.** An optical microscope (Nikon LV150N) was used to examine the morphology and capture images of ML $MoS_2$.

XPS measurements were conducted using a Thermo Fisher Scientific Instruments K Alpha+ XPS system. The X-ray beam had an energy of 1486.6 eV and a beam spot size of 400 μm. The calibration of Mo 3d orbital binding energy was achieved by referencing it to the C1s binding energy, which was measured at 284.8 eV.

Raman and photoluminescence (PL) spectra were measured using a LabRAM HR, Horiba Jovin Yvon spectrometer equipped with a 532 nm laser. The spectra was taken with a 100X objective lens having 0.9 numerical aperture. All optical measurements were done in ambient conditions.

Atomic resolution high-angle annular dark-field scanning transmission electron microscopy (HAADF-STEM) images were acquired with Hitachi HD2700C dedicated STEM with Cs probe corrector at accelerating voltage of 200 kV.

**Device Fabrication and electrical measurements.**

To investigate the electrical transport properties of ML $MoS_2$, electrical contacts were made using the standard photolithography method followed by the thermal vapour deposition of Cr/Au (5nm/65nm) and lift-off in acetone.

Various electrical measurements were done using Keithley 2450 source meter and Keithley 2614B multi-channel source meter unit in a home-built probe station. For vacuum and low-temperature transport measurements, wire-bonded devices were placed inside a Janis close cycle refrigerator (CCR) with a base temperature of 10K and a base pressure of $10^{-6}$ mbar. Transport measurements in low temperatures were recorded using either Keithley 2614B or Yokogawa GS820 multi-channel source measurement unit.

**Sample preparation and STM measurements.**

For sample preparation, we utilized a patterned gold substrate with a gold pad lateral size of 100 microns deposited on a $SiO_2$/Si wafer. We employed a nail polish-based dry transfer technique to transfer the flakes. The bulk graphite flake served as the flat background for scanning ML $MoS_2$. The exfoliated bulk graphite flake, with a lateral size of approximately 70 micrometres, was transferred onto the patterned gold pad, making it easily locatable with the camera. The ML $MoS_2$ was grown using chemical vapour deposition (CVD) and had a large lateral size of about 50 microns. It was then transferred onto the bulk graphite flake. We used the SPI Flash Dry silver colloidal solution to establish contact between the gold pad and the STM sample plate. All STM/STS measurements were conducted at a base temperature of 4.3 K and a pressure of $1.5 \times 10^{-9}$ mBar using the Scienta Omicron Ultra High Vacuum Low Temperature Scanning Tunneling Microscope (UHV-LT-STM). For the measurements, we employed a chemically etched tungsten (W) tip, which was characterized over the Cu(111) surface prior to the measurements. The data obtained were analyzed using SPIP (Image Metrology) 6.0.9 software.

AUTHOR INFORMATION

**Corresponding Author**

**Atikur Rahman**- Department of Physics, Indian Institute of Science Education and Research, Pune, Maharashtra, 411008, India

**Authors**

**Sudipta Majumder-** Department of Physics, Indian Institute of Science Education and Research, Pune, Maharashtra, 411008, India

**Vaibhav Bhalve**- Department of Physics, Indian Institute of Science Education and Research, Pune, Maharashtra, 411008, India

**Rahul Chand**- Department of Physics, Indian Institute of Science Education and Research, Pune, Maharashtra, 411008, India

**Gokul M.A.**- Department of Physics, Indian Institute of Science Education and Research, Pune, Maharashtra, 411008, India

**Sooyeon Hwang**- Center for Functional Nanomaterials (CFN), Brookhaven National Laboratory, U.S. Department of Energy Office.

**G. V. Pavan Kumar**- Department of Physics, Indian Institute of Science Education and Research, Pune, Maharashtra, 411008, India

**Aparna Deshpande**- Department of Physics, Indian Institute of Science Education and Research, Pune, Maharashtra, 411008, India

## Acknowledgement

The authors acknowledge funding support from DST SERB Grant no. C.R.G./2021/005659 and partial funding support under the Indo-French Centre for the Promotion of Advanced Research (CEFIPRA), project no. 6104-2. SM acknowledges CSIR, India, for Senior Research Fellowship (SRF). VB also acknowledges CSIR, India for SRF. Rahul Chand acknowledges the Ministry of Human Resource Development (MHRD) and IISER Pune for SRF. SM acknowledges Sumaiya

Parven, Manisha, and Sayan Maity for valuable discussions. SM acknowledges Rohan Thangaraj for providing Python code to plot the Raman Mapping.

# Probing the Relationship between Defects and Enhanced Mobility in MoS$_2$ Monolayers Grown by Mo Foil

## Supplementary Information

*Sudipta Majumder[1], Vaibhav Walve[1], Rahul Chand[1], Gokul M.A[1], Sooyeon Hwang[1], G. V. Pavan Kumar[1], Aparna Deshpande[1], Atikur Rahman[1]\**

[1]Department of Physics, Indian Institute of Science Education and Research, Pune, Maharashtra, 411008, India

[2]Center for Functional Nanomaterials (CFN), Brookhaven National Laboratory, U.S. Department of Energy Office.

**Contents:**

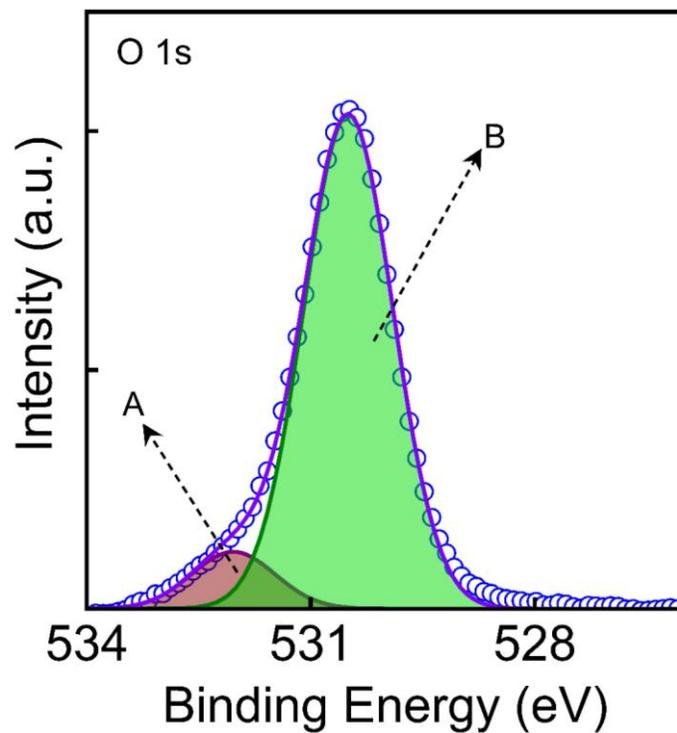

**Figure S1.** XPS spectra of O 1s scan for oxidized Mo foil a) Shows O 1s scan. The O 1s spectrum contains two peaks, A peak at 530.5 eV and B peak at 532eV, corresponding to lattice oxygen and interstitial oxygen.

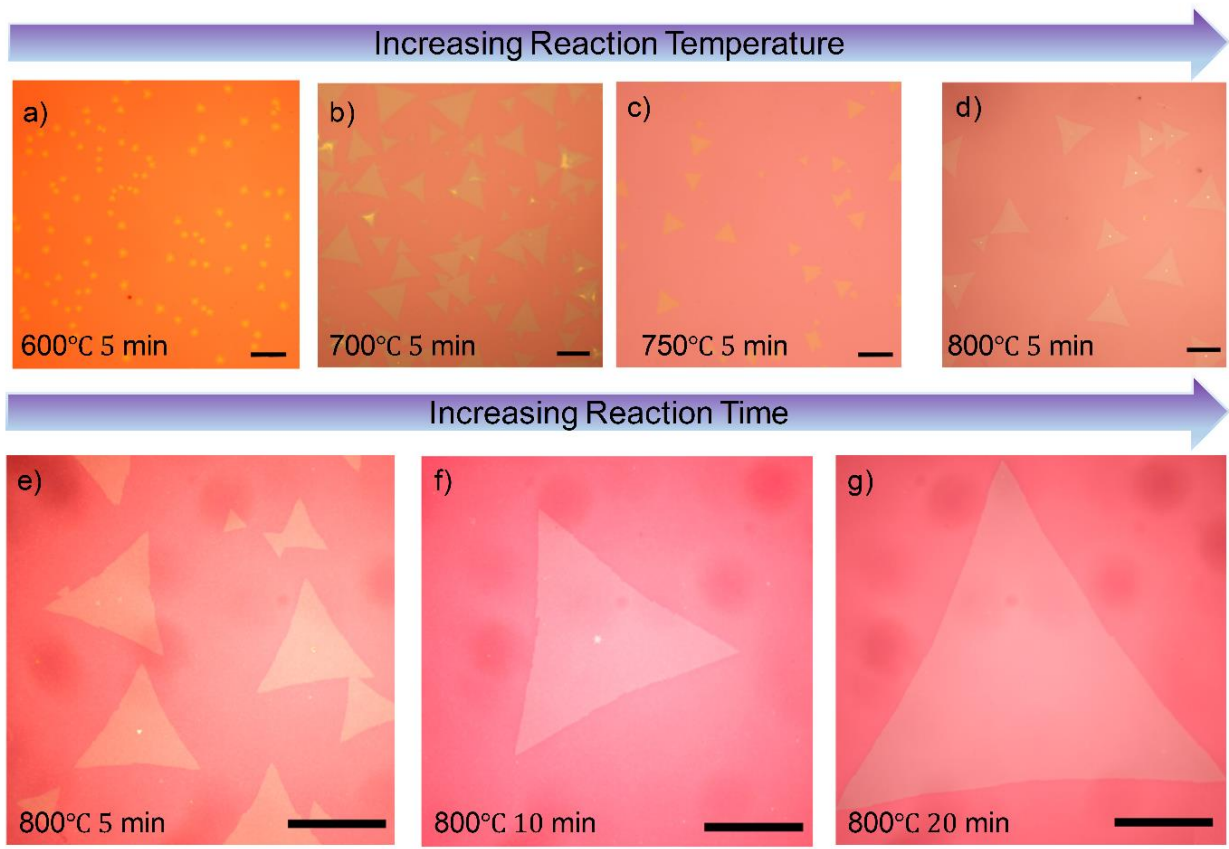

**Figure S2**. Optical pictures of as grown MoS$_2$ in different temperatures and growth times. (a)-(d) the growth of Mo foil assisted MoS$_2$ at different temperatures. The reaction time is 5 minutes. The scale bar is 20 μm. (e)-(g) Optical images for MoS$_2$ growth at different times. The grain domain size is increasing with increase in reaction time indicating planer growth. Scale bar is 20 μm

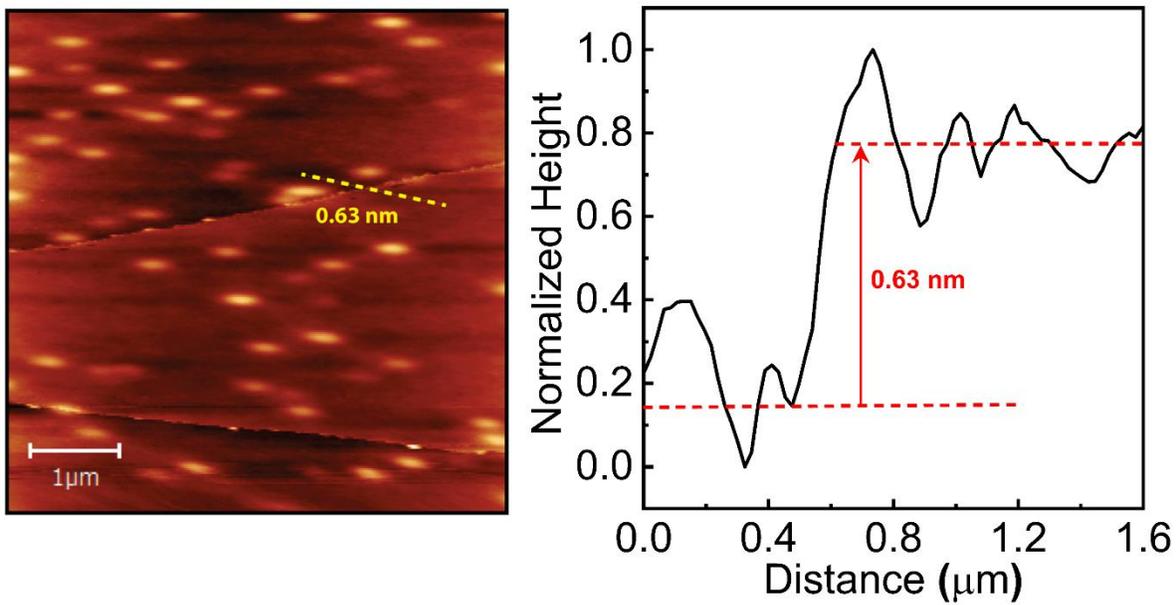

**Figure S3**. Atomic force microscopy image of Mo foil grown MoS$_2$ and its corresponding height profile.

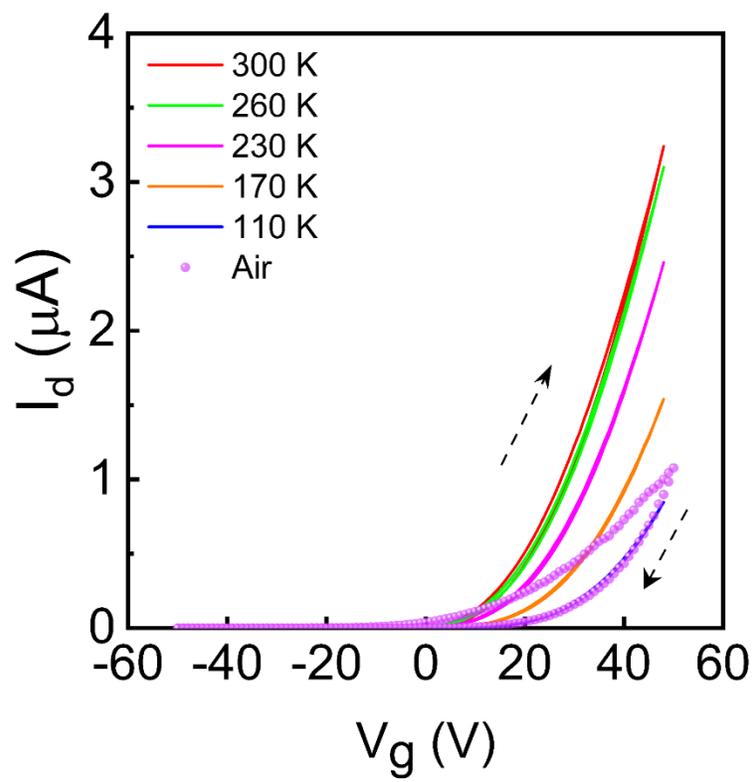

**Figure S4.** Gate hysteresis loop in ambient and vacuum condition at different temperatures. The arrow shows the sweep direction. The vanishing gate hysteresis shows the effect of absorbates.

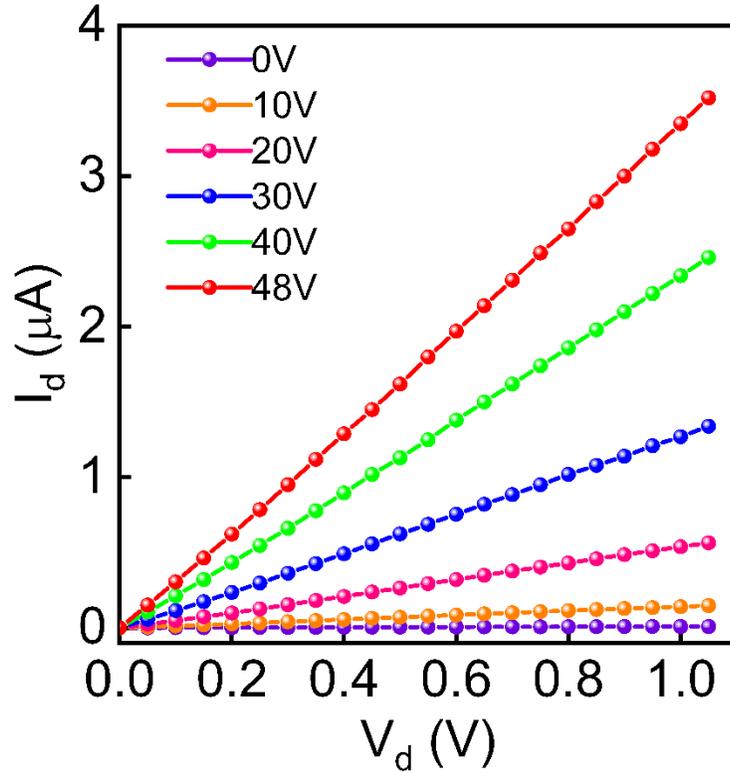

**Figure S5.** Output characteristics of Low mobility device (LM device) in vacuum at different gate voltages. The output characteristics are linear so contacts are ohmic.

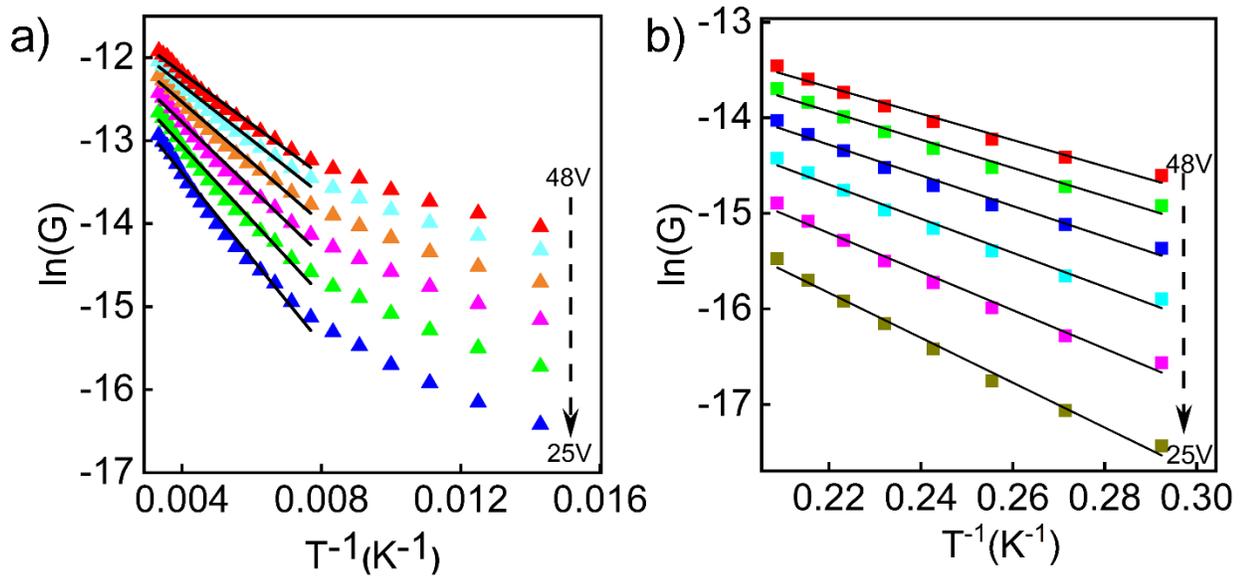

**Figure S6.** Thermally activated transport and Variable Range Hopping (VRH) transport in LM device. a) Logarithm of conductivity (ln G) plotted against temperature (T) for different gate voltages. The graphs are fitted with thermally activated transport model in high temperatures and activation energy is calculated from the slope. b) Mott type VRH for different gate voltages at lower temperatures for the same device.

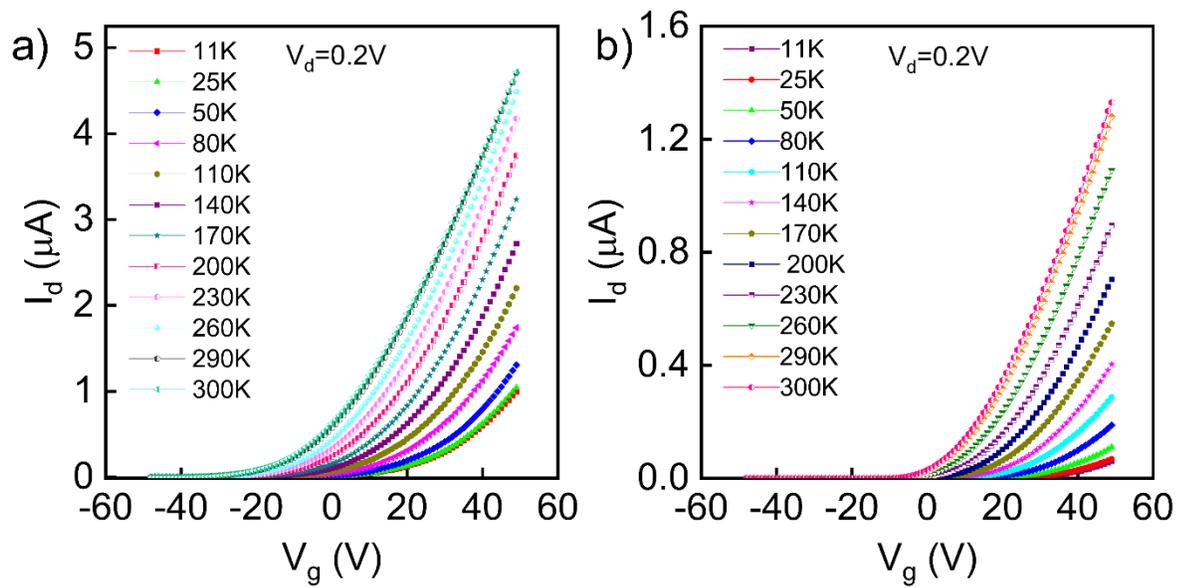

**Figure S7.** The temperature dependent transfer characteristics in (a) High mobility and (b) Low mobility devices at a fixed source drain current 0.2V.

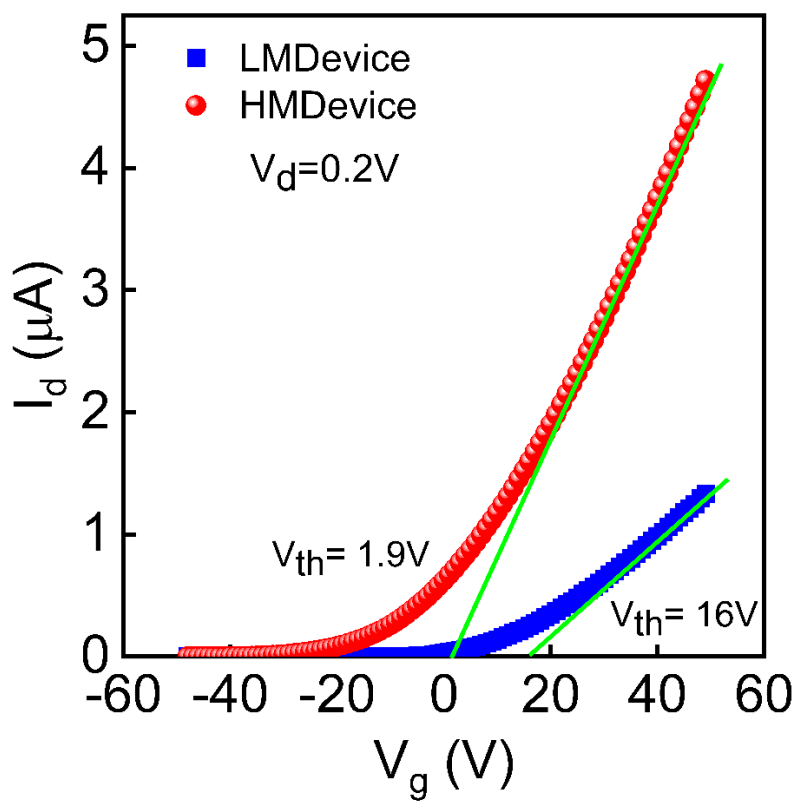

**Figure S8.** (a) Gate dependence of LM device in vacuum conditions. (b) Gate characteristics of LM and HM device. The threshold voltage of HM device is in more negative side that confirms the device is more doped.

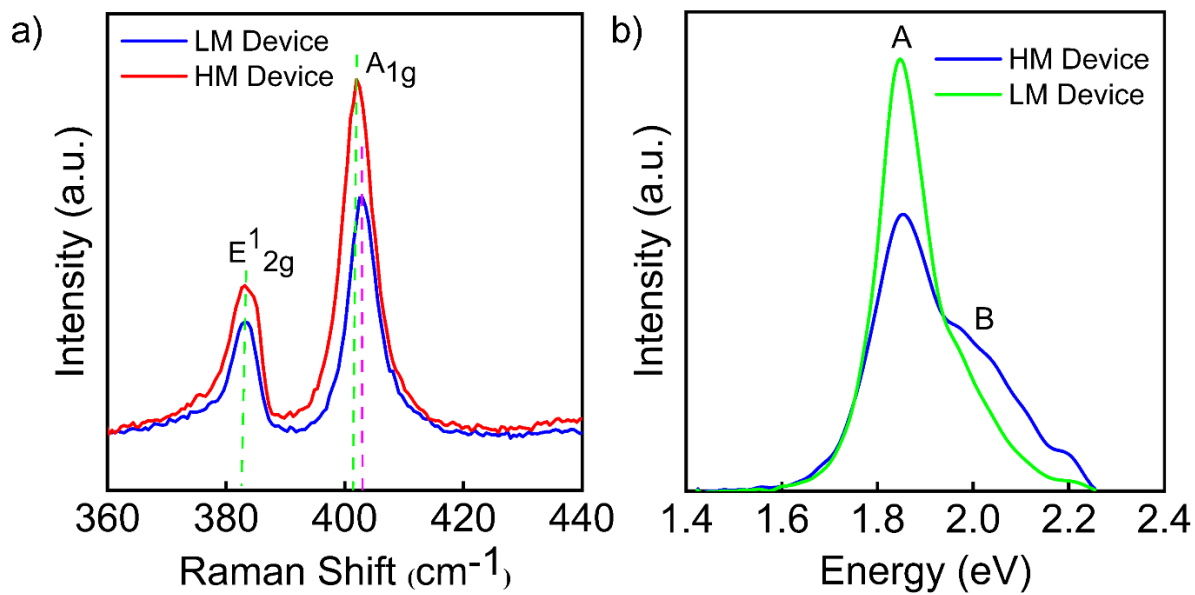

**Figure S9.** a) Raman sprectrum for high and low mobility devices. The $A_{1g}$ peak shows a red shift for HM device indicating relatively higher density of doping. b) The photoluminescence spectra of two devices. B excitonic peak for HM device has a relatively higher intensity than LM device.

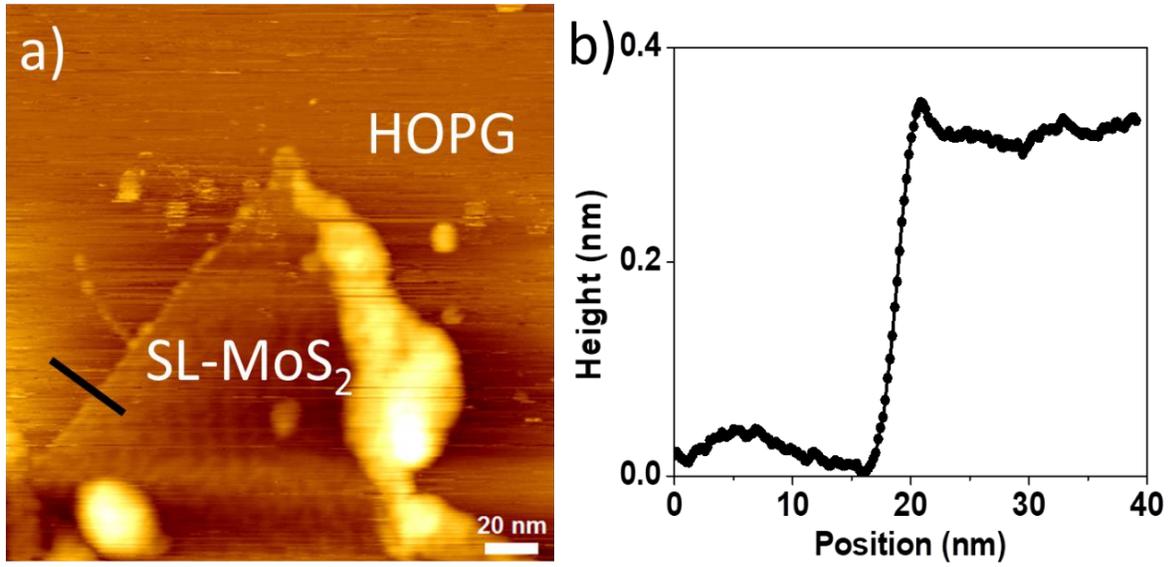

**Figure S10**. a) Large area image showing the STM topography of ML MoS$_2$ flake over HOPG. (VB=2 V, Is=100 pA) b) Step Height profile of MoS$_2$ over HOPG over the region shown in the (a).

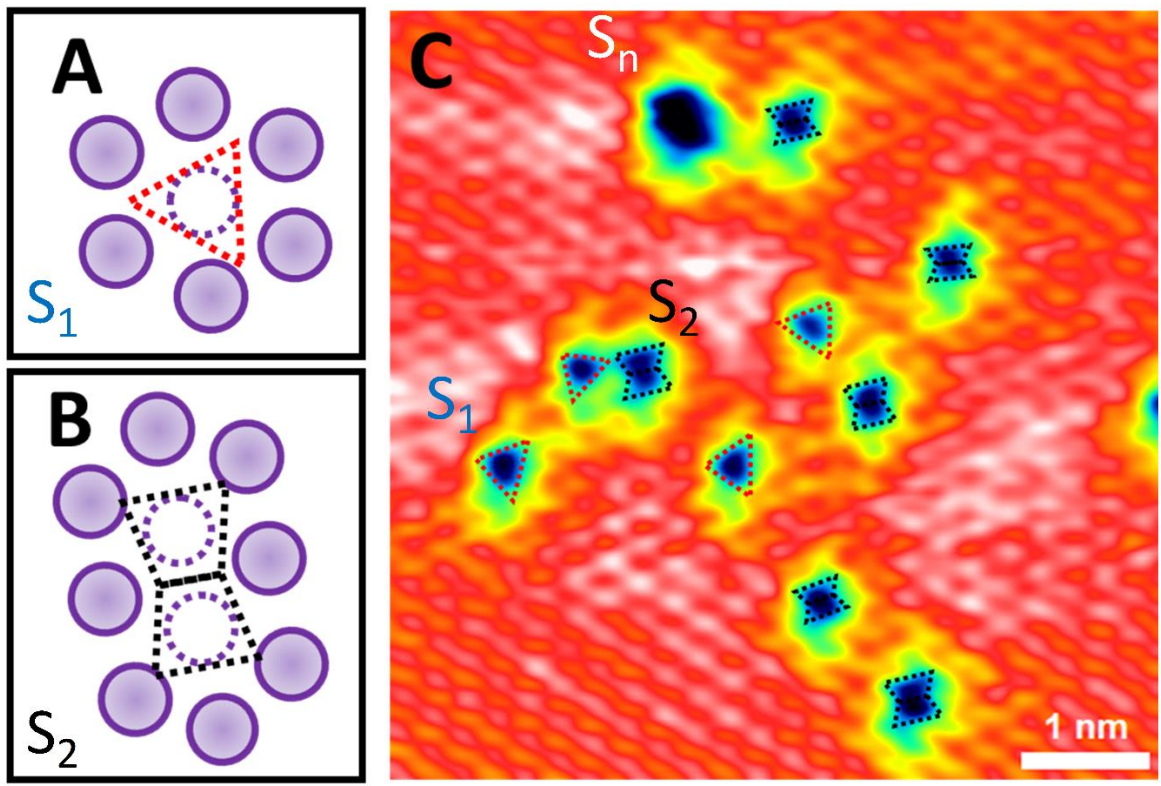

**Figure S11.** Characterization of $S_1$ and $S_2$ vacancies: a) The triangular shape of the single Sulfur defect ($S_1$) allowed by symmetry. b) Hourglass-like shape of double sulfur vacancies ($S_2$). c) The different $S_1$ (red) and $S_2$ (black) defects are shown on the STM topography. ($V_B=1$ V, $I_s=150$ pA)

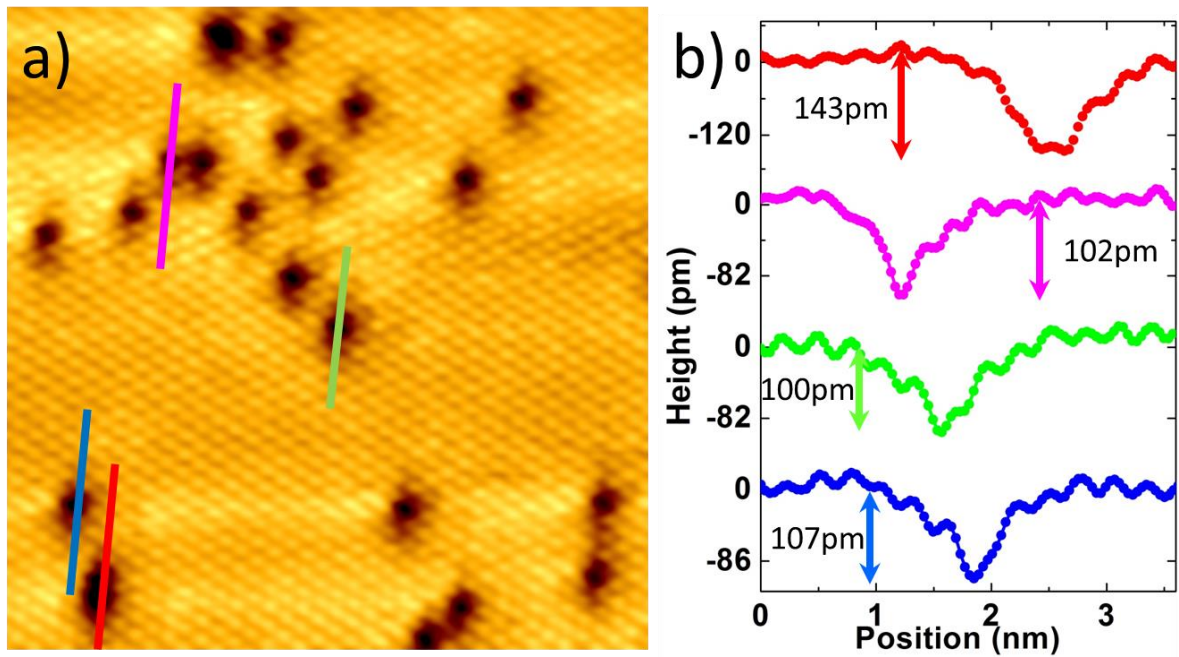

**Figure S12.** a) STM topography of the ML MoS$_2$ (VB=1 V, Is=150 pA) b) Line profiles taken over the different defects in the STM topography and corresponding depth in pm at each defect site are marked on the spectra.